\pdfoutput=1 
\documentclass{JINST}

\title{Run 2 Upgrades to the CMS Level-1 Calorimeter Trigger}

\author{
B. Kreis$^a$\thanks{Corresponding author.}, J. Berryhill$^a$, R. Cavanaugh$^{a,b}$, K. Mishra$^a$,  R. Rivera$^a$, L. Uplegger$^a$, 
L. Apanasevich$^b$, J. Zhang$^b$, 
J. Marrouche$^c$, N. Wardle$^c$, 
R. Aggleton$^d$, F. Ball$^d$, J. Brooke$^d$, D. Newbold$^d$, S. Paramesvaran$^d$, D. Smith$^d$, 
M. Baber$^e$, A. Bundock$^e$, M. Citron$^e$, A. Elwood$^e$, G. Hall$^e$, G. Iles$^e$, C. Laner$^e$, B. Penning$^e$, A. Rose$^e$, A. Tapper$^e$, 
C. Foudas$^f$, 
F. Beaudette$^g$,  L. Cadamuro$^g$, L. Mastrolorenzo$^g$, T. Romanteau$^g$, J. B. Sauvan$^g$, T. Strebler$^g$, A. Zabi$^g$, 
R. Barbieri$^h$, I. A. Cali$^h$, G. M. Innocenti$^h$, Y.-J. Lee$^h$, C. Roland$^h$, B. Wyslouch$^h$, 
M. Guilbaud$^i$, W. Li$^i$, M. Northup$^i$, B. Tran$^i$, 
T. Durkin$^j$, K. Harder$^j$, S. Harper$^j$, C. Shepherd-Themistocleous$^j$, A. Thea$^j$, T. Williams$^j$, 
M. Cepeda$^k$, S. Dasu$^k$, L. Dodd$^k$, R. Forbes$^k$, T. Gorski$^k$, P. Klabbers$^k$, A. Levine$^k$, I. Ojalvo$^k$, T. Ruggles$^k$,  N. Smith$^k$, W. Smith$^k$, A. Svetek$^k$, J. Tikalsky$^k$, and M. Vicente$^k$\\
\llap{$^a$}Fermi National Accelerator Laboratory,
Batavia, Illinois, U.S.A.\\
\llap{$^b$}University of Illinois at Chicago,
Chicago, Illinois, U.S.A.\\
\llap{$^c$}CERN, European Organization for Nuclear Research,
Geneva, Switzerland\\
\llap{$^d$}University of Bristol,
Bristol, United Kingdom\\ 
\llap{$^e$}Imperial College, University of London, 
London, United Kingdom\\
\llap{$^f$}University of Io\'{a}nnina,
 Io\'{a}nnina, Greece\\
\llap{$^g$}Laboratoire Leprince-Ringuet, Ecole Polytechnique,
IN2P3-CNRS, Palaiseau, France\\
\llap{$^h$}Massachusetts Institute of Technology, 
Cambridge, Massachusetts, U.S.A.\\
\llap{$^i$}Rice University,
Houston, Texas, U.S.A.\\
\llap{$^j$}Rutherford Appleton Laboratory,
Didcot, United Kingdom\\
\llap{$^k$}University of Wisconsin,
Madison, Wisconsin, U.S.A.\\
E-mail: \email{kreis@fnal.gov}
}

\abstract{The CMS Level-1 calorimeter trigger is being upgraded in two stages to maintain performance as the LHC increases pile-up and instantaneous luminosity in its second run. 
In the first stage, improved algorithms including event-by-event pile-up corrections are used.
New algorithms for heavy ion running have also been developed. 
In the second stage, higher granularity inputs and a time-multiplexed approach allow for improved position and energy resolution. 
Data processing in both stages of the upgrade is performed with new, Xilinx Virtex-7 based AMC cards.}


\usepackage{lineno}
\usepackage{subfig}

\begin{document}


\section{Introduction}\label{sec:intro}
Following Long Shutdown 1, the second run of the Large Hadron Collider (LHC) at CERN is now underway with an increased center-of-mass energy of 13 TeV \cite{lhc}.
In Run 2, the LHC will operate with a bunch spacing of 25 ns and beam parameters that exceed design performance.
Instantaneous luminosity is expected to reach $\sim$$1.5 \times 10^{34}$ cm$^{-2}$s$^{-1}$ early in Run 2 and increase further, 
and the number of simultaneous inelastic collisions per crossing, or pile-up, is expected to reach $\sim$50, both well above design.
Instantaneous luminosity in heavy ion collisions is also expected to increase by a factor of between four and eight \cite{tdr}.

The Level-1 (L1) trigger at CMS, which uses dedicated readout paths from the calorimeter and muon detectors to trigger the full readout of CMS, must be upgraded 
to maintain acceptance for proton and heavy ion collision events of interest without exceeding the 100 kHz limit \cite{tdr}.       
At the instantaneous luminosity expected in 2015, using the same trigger thresholds as at the end of Run 1 without an upgrade would lead to trigger rates roughly six times the limit.
Upgrades to the L1 calorimeter trigger, the part of the L1 trigger processing data from the calorimeter detectors, are described here. 
The L1 calorimeter trigger finds the highest transverse energy jet, tau, and electron/photon candidates and computes global energy sums.
Electromagnetic calorimeter (ECAL), hadronic calorimeter (HCAL), and hadronic forward calorimeter (HF) trigger towers 
provide transverse energies with reduced energy and position resolution, called trigger primitives.  
The position resolution is set by the size of the trigger towers.
The granularity corresponds to seventy-two divisions in azimuthal angle ($\phi$) 
and fifty-six divisions in the pseudorapidity ($\eta$) range instrumented with the ECAL and HCAL, $|\eta|<3.0$. 
HF trigger towers provide an additional four divisions in both positive and negative $\eta$ (soon twelve, as described in section~\ref{sec:stage2-hw}).
In Run 1, the trigger primitives were processed by the Regional Calorimeter Trigger (RCT) \cite{run1trig}.  
The resulting regional data were then processed by the Global Calorimeter Trigger (GCT) whose output was sent to the Global Trigger (GT), where the L1 trigger decision was made.

The L1 calorimeter trigger is being upgraded in two stages.  
The first stage, Stage 1, was a partial upgrade that went online in 2015.  
As described section~\ref{sec:stage1}, 
the GCT was replaced with a new data processing card capable of executing improved algorithms,
including event-by-event pile-up subtraction and dedicated algorithms for heavy ion running.
In 2016, the second stage, Stage 2, will go online.
As described in section~\ref{sec:stage2}, in this upgrade, whole events are time multiplexed and analyzed at the global level at full trigger tower granularity.

\section{Stage 1 Upgrade}\label{sec:stage1}
The Stage 1 calorimeter trigger is a partial upgrade on the way to the full Stage 2 upgrade.
The GCT was replaced by a Master Processor, Virtex-7 (MP7) AMC card \cite{mp7}.  
This is the first application of this card at CMS.  Several MP7s will be used in the Stage 2 upgrade.
Data communication was also partially upgraded.
The ECAL was retrofitted with new optical links that will also be used in the Stage 2 trigger,
and new electronics are used to duplicate the RCT output to the MP7. 

\subsection{New Optical Links}\label{sec:stage1_links}

Data communication from the ECAL was upgraded from electrical to optical by retrofitting the ECAL Trigger Concentrator Cards \cite{tcc} with Optical Synchronization and Link Boards (oSLBs) \cite{oSLB}.  These mezzanine boards synchronize the ECAL trigger primitives from up to eight trigger towers at the LHC bunch crossing frequency and concentrate them onto 4.8 Gbps links.  One copy is transmitted to the RCT, and a second is transmitted to the Stage 2 trigger to allow for parallel running.  There are 576 oSLBs in total.

The RCT receives the ECAL trigger primitives with new Optical Receiver Mezzanines (oRMs) \cite{oSLB}.  There are 504 oRMs, where seventy-two of them operate with two receivers.  On the output side, RCT is retrofitted with eighteen Optical Regional Summary Cards (oRSCs) \cite{oRSC}.  Each oRSC transmits one copy of an RCT crate's output to the GCT via eleven 2 Gbps links and up to six copies on pairs of 10 Gbps links.  Duplicating the data from RCT allows systems being commissioned to run in parallel for testing.  It also allows for the downstream data processing to be expanded by connecting multiple MP7s, although there are no plans for this.

\subsection{MP7 Data Processing Card}\label{sec:stage1_card}

The upgraded data processing in the Stage 1 trigger occurs on a single MP7 card 
featuring a Xilinx Virtex-7 XC7VX690T FPGA \cite{mp7}.  
The data throughput and computational power of this card are sufficient to not only match but surpass  
the performance of the GCT, which contains over twenty older FPGAs distributed across multiple boards.
Approximately twenty percent of the flip flops, forty-five percent of the LUTs, and forty percent of the block RAMs are used in the proton-proton algorithm firmware.

The MP7 has seventy-two input and seventy-two output optical links that can operate at up to 10 Gbps.
In the Stage 1 trigger, thirty-six 10 Gbps links receive data from the RCT, 
and fourteen 3 Gbps transceivers send data to the GT.
The card is housed in a Vadatech VT892 MicroTCA crate, and
additional serial and LVDS electrical I/O occurs via the backplane.
An AMC13 card \cite{amc13} provides clock and timing signals and the L1 trigger decision via LVDS.
For every triggered event, the AMC13 reads out a copy of the MP7's input and output via a 5 Gbps serial link.
The inputs (outputs) for the two bunch crossings before and after are also read out for approximately one percent (all) of triggered events. 
These data are sent to the CMS data acquisition system and are available for monitoring and offline analysis. 
Data to configure the MP7's lookup tables (LUTs) and registers are sent via Ethernet to a NAT-MCH MicroTCA Carrier Hub
and then on to a MP7 serial link following the IPbus protocol \cite{ipbus}.

\subsection{Improved Algorithms}\label{sec:stage1_algo}

In proton-proton collision running, the Stage 1 MP7 finds the four highest transverse energy jet, tau, and electron/photon candidates and computes global energy sums using the regional data from the RCT.  The improvement crucial to handling the increased instantaneous luminosity and pile-up from the LHC is the subtraction of an event-by-event estimate for the energy resulting from pile-up before the outputs are computed.  This is described in section~\ref{sec:stage1_pu}.  In section~\ref{sec:stage1_taus}, improvements to the tau trigger are highlighted.  Section~\ref{sec:stage1_hi} describes the all-new suite of algorithms that have been developed to cope with the increased instantaneous luminosity expected in heavy ion collision running.  All of the algorithms are pipelined to accept a new event every 25 ns.

The inputs to the Stage 1 MP7 provided by the RCT are the total ECAL plus HCAL transverse energy in $4\times4$ regions of trigger towers, the transverse energy of single HF trigger towers (also called regions here), and isolated and non-isolated electron/photon candidates formed from $2\times1$ combinations of ECAL trigger towers.  The regions form a grid with 22 $\eta$ slices and 18 $\phi$ slices.  The electron/photon candidates arrive nine bunch crossings after the regions' transverse energies.  As a result, the Stage 1 MP7 is allotted approximately twenty bunch crossings of latency for the jet, tau, and global energy sum algorithms and nine less for the electron/photon algorithm.

\subsubsection{Pile-up Subtraction}\label{sec:stage1_pu}

\begin{figure}[tbp]%
    \centering
    \subfloat[]{{\includegraphics[width=.4\textwidth]{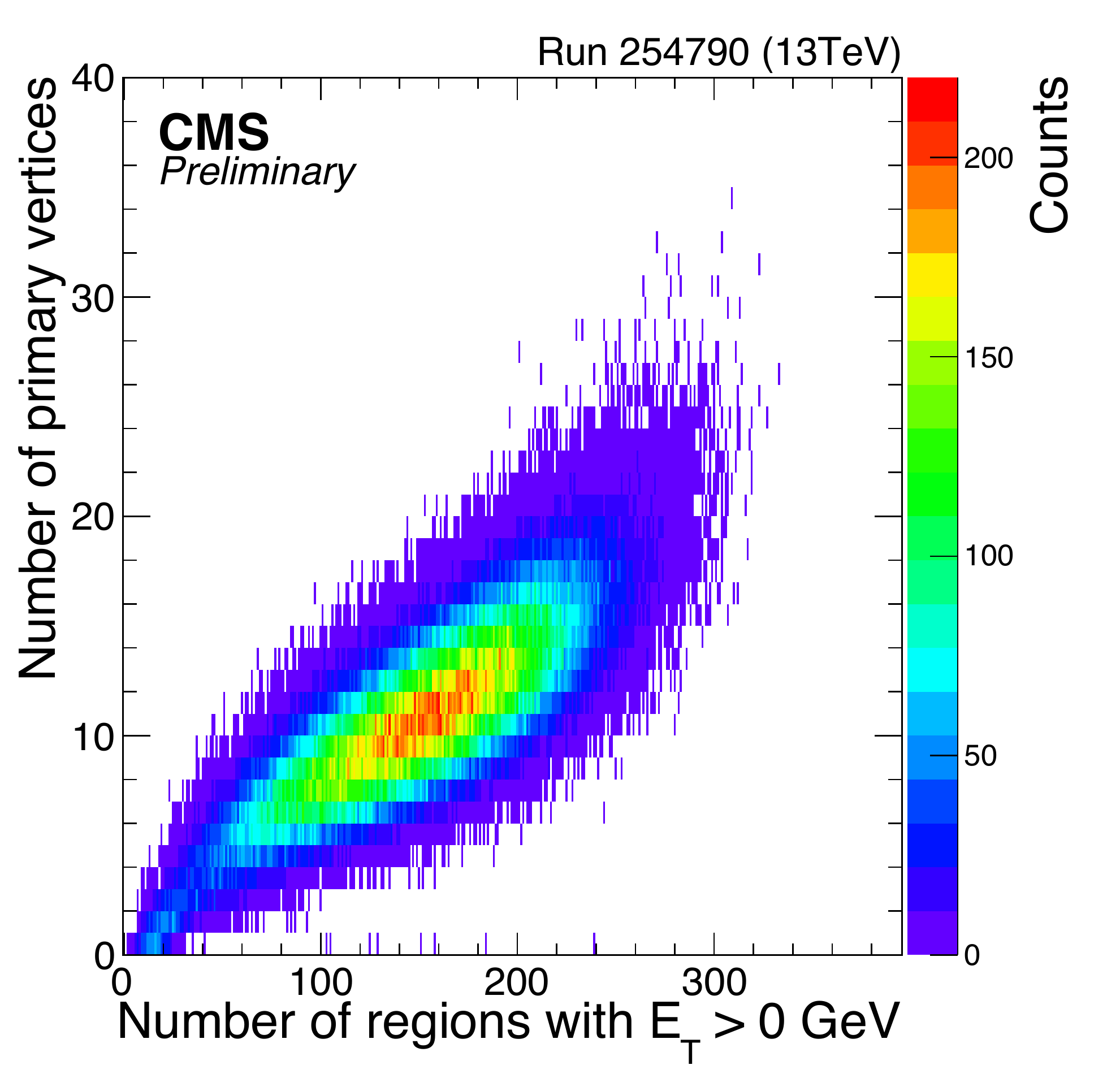} }\label{fig:pum0}}%
    \qquad
    \subfloat[]{{\includegraphics[width=.4\textwidth]{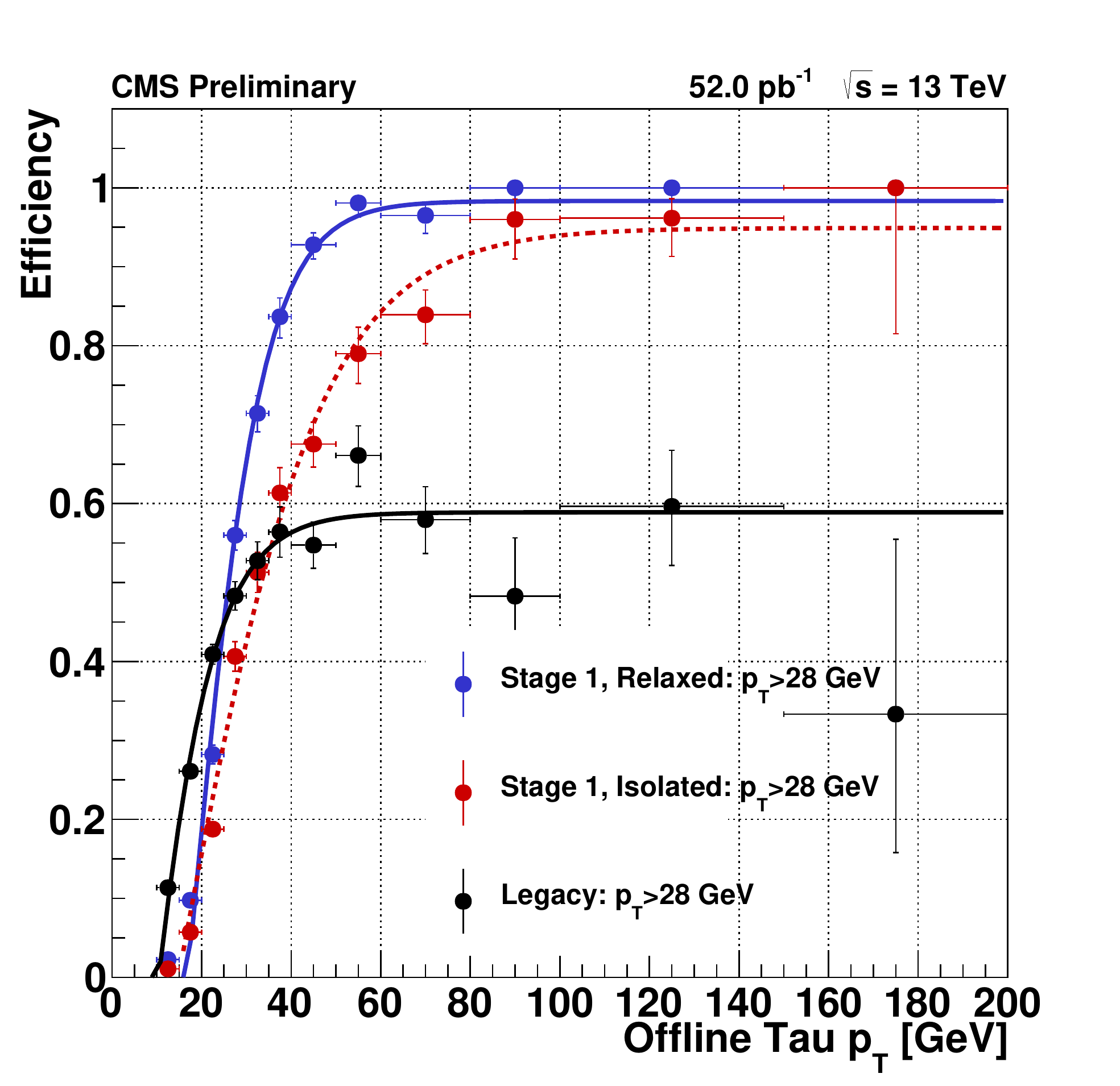} }\label{fig:taus}}%
    \caption{(a) Correlation between the number of regions with non-zero transverse energy ($\mathrm{E}_{\mathrm{T}}$) and the number of reconstructed primary vertices in Run 2 data. 
(b) Isolated and relaxed (no isolation requirement) tau trigger efficiency as a function of offline 
transverse momentum ($\mathrm{p}_{\mathrm{T}}$) for an online requirement of $\mathrm{p}_{\mathrm{T}}>28$ GeV.
The Run 1, or legacy, efficiency for taus is shown for comparison \cite{hlt_tau}.  Note that in Run 1, the di-tau High Level Trigger was seeded with the logical OR of a L1 di-tau requirement and L1 di-jet requirement to improve efficiency.   Events are required to contain a loosely selected offline tau. }%
    \label{fig:example}%
\end{figure}

The number of regions with non-zero transverse energy is used as a global, indirect estimator for pile-up.
The correlation between this number and the number of reconstructed primary vertices is shown in figure~\ref{fig:pum0}.
The number is converted to the pile-up energy to subtract from each region with LUTs.
Because pile-up energy density depends on $\eta$, each $\eta$ slice has a unique LUT.

Pile-up subtraction improves the performance of all the Stage 1 proton-proton algorithms.  Jet and tau candidates and global energy sums are computed from the pile-up subtracted regions.  The pile-up subtracted regions are also used to determine if tau and electron/photon are isolated.  One consequence of using a global quantity as an estimator of pile-up is that the whole event must be read in before the subtraction can be done and the subsequent algorithms can proceed.  This drove the choice to do minimal pipelining, with most algorithms operating on half the detector in parallel with an 80 MHz clock.   

\subsubsection{Tau Algorithms}\label{sec:stage1_taus}

The Stage 1 upgrade brings two improvements to tau triggers.  The first is that the feature size of tau candidates was reduced from a $3\times3$ square of regions to $2\times1$.  As is shown in figure~\ref{fig:taus}, changing to this more appropriate size improves the efficiency by approximately forty percent.  The second improvement is that the Stage 1 trigger provides isolated tau candidates to the GT for the first time.  The isolation requirement leads to a large drop in rate with only a small reduction in efficiency (shown in figure~\ref{fig:taus}).  The isolation decision is based on the the relative isolation, $({\mathrm{E}_{\mathrm{T}}}_{3\times3} - {\mathrm{E}_{\mathrm{T}}}_{\mathrm{tau}} )/{\mathrm{E}_{\mathrm{T}}}_{\mathrm{tau}}$, where ${\mathrm{E}_{\mathrm{T}}}_{\mathrm{tau}}$ is the transverse energy of the tau candidate and ${\mathrm{E}_{\mathrm{T}}}_{3\times3}$ is the transverse energy of the $3\times3$ square of regions surrounding the highest energy region of the tau candidate.  LUTs addressed with the two transverse energies are used to compute the relative isolation decision.  The four highest transverse energies of the isolated tau candidates are sent to the GT with a coarse resolution using bits originally reserved for sums of transverse energy in the HF.  All tau candidates can have an $\eta$-dependent transverse energy correction applied via a LUT. 

\subsubsection{Heavy Ion Algorithms}\label{sec:stage1_hi}

A suite of algorithms has been developed to cope with the increased instantaneous luminosity expected in heavy ion collision running.  Similar to the pile-up subtraction described in section~\ref{sec:stage1_pu}, background is subtracted from the regions before computing all outputs (except the global energy sums in this case).  The background subtracted from each $\eta$ slice is the mean transverse energy of the $\eta$ slice.  

\begin{figure}[tbp]%
    \centering
    \subfloat[]{{\includegraphics[width=.5\textwidth]{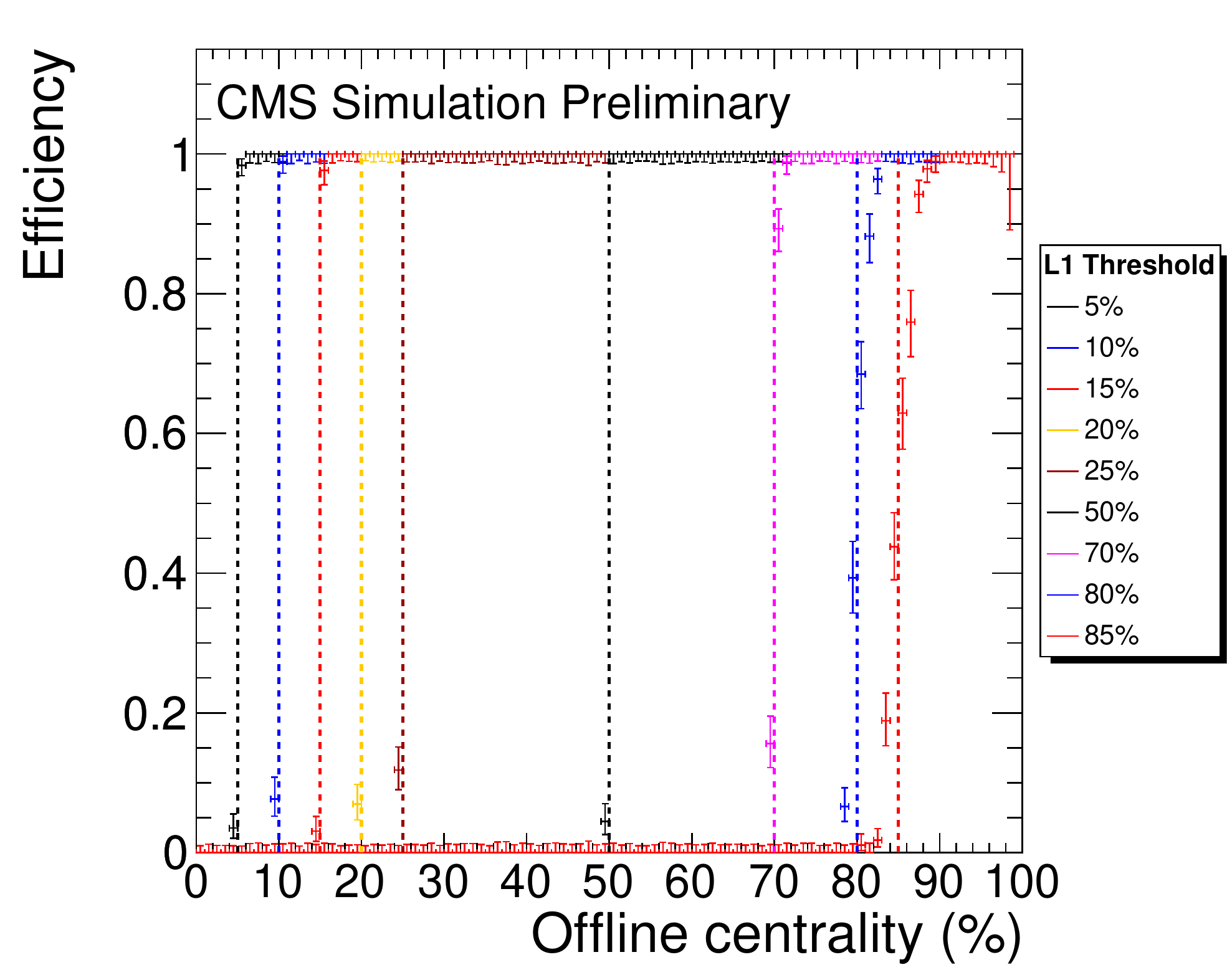} }\label{fig:centrality}}%
    \qquad
    \subfloat[]{{\includegraphics[width=.4\textwidth]{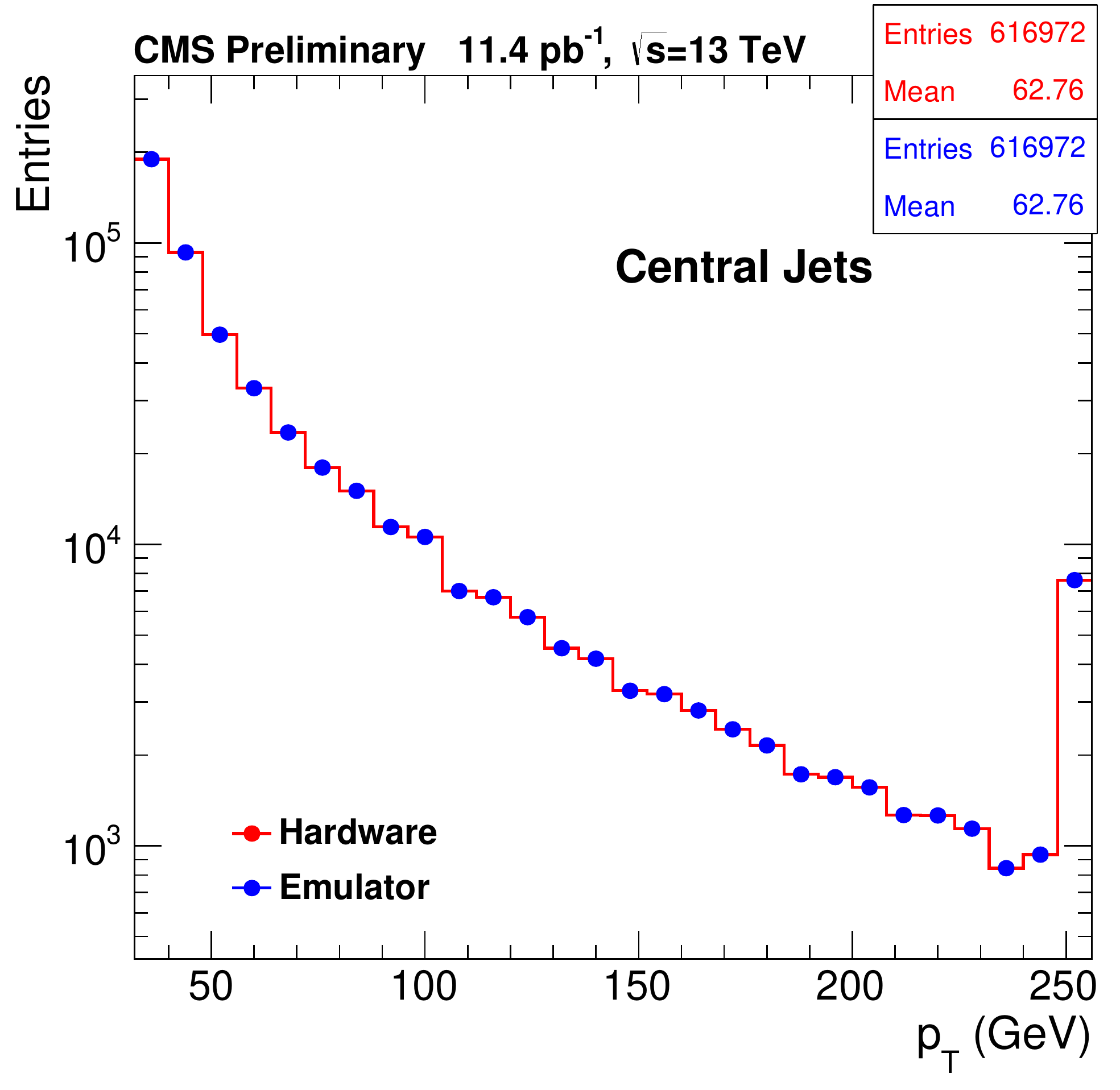} }\label{fig:stage1_emulator}}%
    \caption{(a) Efficiency of centrality trigger as a function of offline centrality for a range of thresholds in simulated Pb-Pb collision events with $\sqrt{s}=5.02$ TeV. 
    (b) Perfect, bit-level agreement on central ($|\eta|<3.0$) jet candidate transverse energy, labelled $\mathrm{p}_{\mathrm{T}}$ here, between the Stage 1 MP7 (hardware) and C++ emulator. }%
\end{figure}

Several other changes with respect to the proton-proton algorithms are made for heavy ion running.
Electron/photon candidates are split by their location in the ECAL  
-- barrel ($|\eta|<1.479$) versus endcap ($1.479<|\eta|<3.0$) -- instead of by isolation.
The transverse energy of a jet candidate is taken as the largest $2\times2$ transverse energy inside the $3\times3$ jet candidate.    
The tau candidate algorithm is repurposed to find the highest energy regions as a seed for a single track High Level Trigger.
Finally, the total transverse energy in the HF is used as an estimator for centrality, which is output in place of the isolated tau candidates.  The expected performance of the centrality trigger is shown in figure~\ref{fig:centrality}.

\subsection{Commissioning and Running}\label{sec:stage1_running}

The duplication of RCT output provided by the oRSCs allowed the Stage 1 trigger to run
in parallel with the GCT for a commissioning phase early in Run 2.  
In this setup, the output of the Stage 1 MP7 was sent to a test version of the GT.  
In August 2015, the Stage 1 MP7 was connected to the production GT in place of the GCT.
A second MP7 is currently being used to commission the heavy ion firmware with a copy of the RCT output.

The Stage 1 MP7 is controlled and monitored by online software integrated into the central L1 trigger software \cite{software}.
At a rate of approximately 100 Hz, the copy of Stage 1 MP7 inputs and outputs read by the AMC13 is used by a data quality monitoring system \cite{dqm} to compare the outputs with a bit-level emulation of the firmware implemented in C++.    
This comparison is monitored in real time by the shift crew at CMS.  
Perfect agreement on jet candidate transverse energy is shown in figure~\ref{fig:stage1_emulator}.

\section{Stage 2 Upgrade}\label{sec:stage2}

The Stage-2 trigger finds the twelve highest transverse energy jet, tau, and electron/photon candidates and computes global energy sums with algorithms operating on the calorimeter detectors' full field of view at trigger tower granularity. 
A time-multiplexed approach makes this possible.
The improved energy and position resolution provides the background rejection required to cope with 
the expected increase in instantaneous luminosity and pile-up.
The Stage-2 trigger is currently being commissioned to begin running in 2016. 
The upgrade will provide CMS with a calorimeter trigger with outstanding performance until Long Shutdown 3 \cite{lhc}.

\subsection{Time Multiplexing Architecture}\label{sec:stage2_arch}

\begin{figure}[tbp] 
\centering
\includegraphics[width=.6\textwidth]{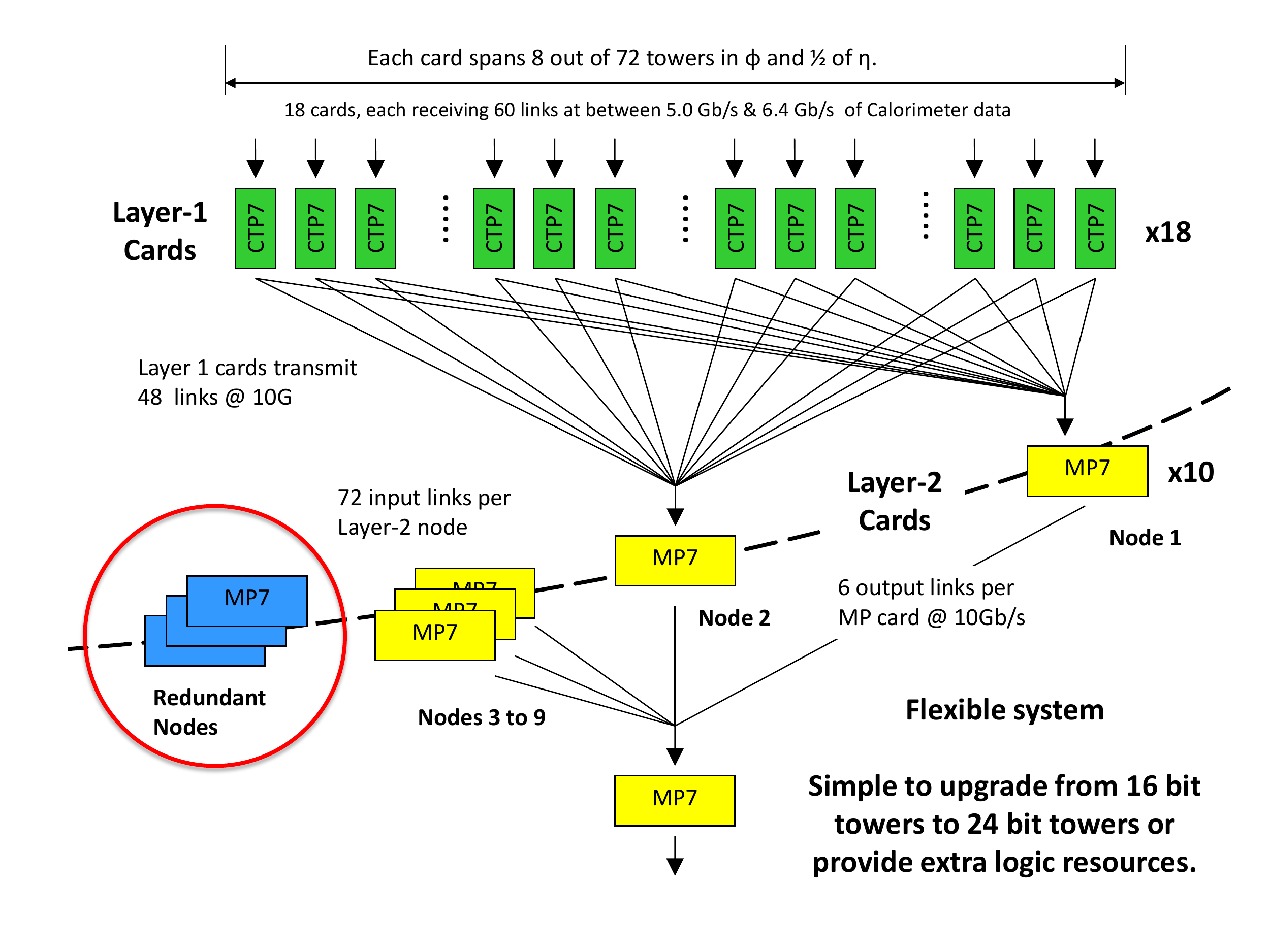}
\caption{Stage-2 trigger architecture.}
\label{fig:stage2}
\end{figure}

The time multiplexing architecture of the Stage-2 trigger is shown in figure~\ref{fig:stage2}.
The trigger is composed of two processing layers.
The first layer, Layer-1, performs pre-processing and data formatting.
The outputs of the Layer-1 pre-processors corresponding to one event are transmitted to 
single processing nodes in the second layer, Layer-2.
The Layer-2 nodes find particle candidates and compute global energy sums.  
These are sent to a demultiplexer board, also an MP7, that formats the data for the upgraded Global Trigger \cite{tdr}.

Both layers are instrumented with Xilinx Virtex-7 FPGAs
on AMC cards conforming to the MicroTCA standard. 
Layer-1 uses CTP7 cards \cite{ctp7} with regional views of the calorimeter detectors.
This can be extended with connections to the MicroTCA backplane allowing for data sharing
between different regions.  Pre-processing only requiring reduced views of the detector,
such as summing ECAL and HCAL transverse energies, occur on the CTP7s.  
Trigger tower position granularity is preserved in this layer.
MP7 cards, described in section~\ref{sec:stage1_card}, form Layer 2.
Each MP7 has access to a whole event at trigger tower granularity. 

Because both the volume of incoming data and
the algorithm latency are fixed, the position of all data within the
system is fully deterministic and no complex scheduling mechanism is
required. The benefit of time multiplexing is extra latency to implement
complex algorithms.  The algorithms are fully pipelined and start processing as
soon as the minimum amount of data is received. 
A total of eighteen CPT7s and ten MP7s (nine as Layer-2 processors and one as the demultiplexer)
are required to implement the trigger to be used starting in 2016.
This can be expanded as necessary.

As in the Stage-1 upgrade, AMC13 cards distribute clock and timing signals and the L1 trigger decision
and perform readout to the CMS data acquisition system.

\subsection{Data Flow}\label{sec:stage2-hw}

The ECAL, HCAL and HF inputs to the Stage-2 trigger are transmitted via new optical links.
The oSLBs described in section~\ref{sec:stage1_links} send a copy of the ECAL trigger primitives
through an optical patch panel 
to Layer-1 on 4.8 Gbps links.  
The HCAL and HF inputs are also sent from upgraded trigger primitive generators at 6.4 Gbps \cite{tdr}.
The HF granularity is increased to twelve divisions in positive and negative $\eta$, 
and other improvements to the trigger primitives, including improved energy resolution, can be made in the future.

The time multiplexing route from Layer-1 to Layer-2 is provided by a Molex FlexPlane patch panel.  
Seventy-two to seventy-two 12-fiber MPO cables are routed in three pizza-box sized enclosures. 
This novel technology may be useful in future LHC electronics systems because it can massively simplify and shrink fiber installations at no extra cost. 

From Layer-1 to Layer-2
a dedicated 16-bit word format has been adopted to encode the sum of ECAL and HCAL transverse energies
along with extra words to compute them separately and extract their ratio (this may be increased to 24-bit words in the future).  
Each $\eta$ slice of seventy-two trigger towers is read out starting from the center of the detector, alternating between the positive and negative side. 
Approximately seven bunch crossings are required to
receive all data from Layer-1.
An input pipeline is necessary to process the data at the incoming rate starting on the
reception of the first data word. For the 32 bits received on each
link, the internal computing frequency achieved is 240 MHz.

\subsection{Algorithm and Commissioning Status}\label{sec:stage2-com}

\begin{figure}[tbp]%
    \centering
    \subfloat[]{{\includegraphics[width=.48\textwidth]{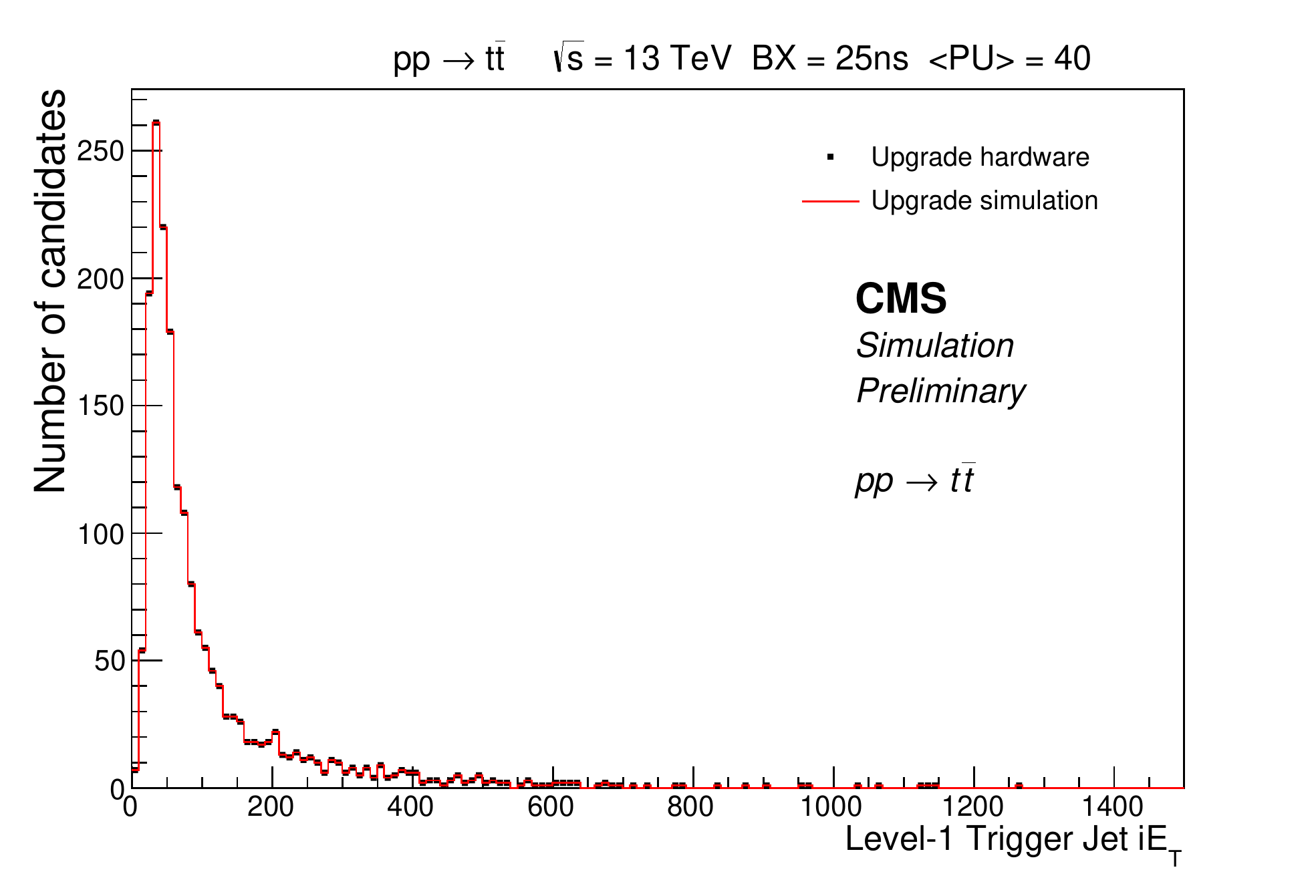} }}%
    \qquad
    \subfloat[]{{\includegraphics[width=.34\textwidth]{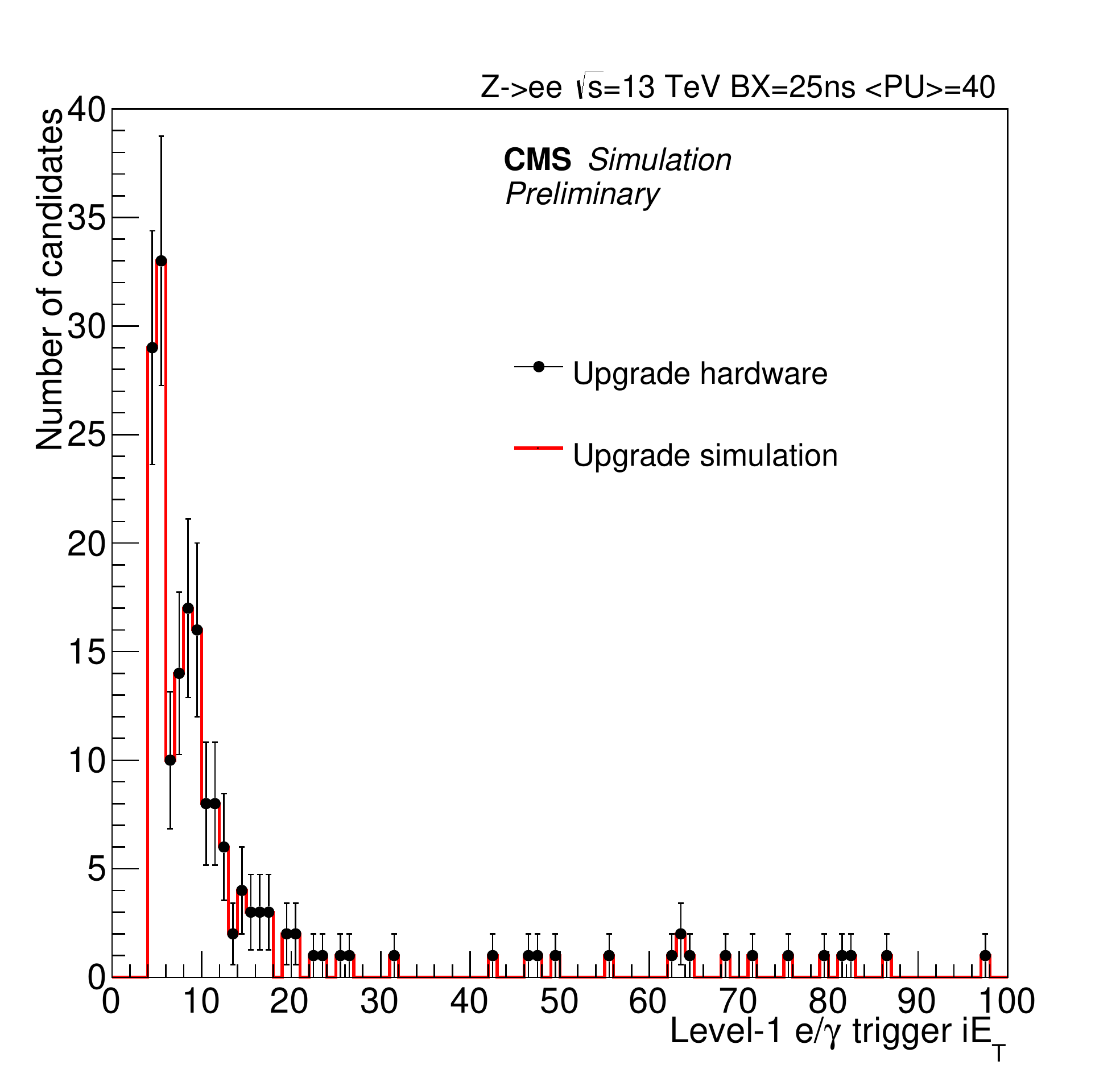} }}%
    \caption{Excellent agreement between Stage-2 firmware and bit-level emulation in C++ in (a) jet transverse energy (i$\mathrm{E}_{\mathrm{T}}$) and (b) electron/photon (e/$\gamma$) transverse energy. }%
    \label{fig:stage2_emfw}%
\end{figure}

The Layer-2 MP7s process whole events at full trigger tower granularity. 
The energy and position resolution of jet, tau and electron/photon candidates and global energy sums 
greatly benefit from this enhanced granularity~\cite{alex}. 
Pile-up estimation is based on trigger tower multiplicity in the case of tau and electron/photon candidates 
and on nearby energy deposits in the case of jets.
Excellent agreement between the firmware and bit-level emulation in C++ is shown in figure~\ref{fig:stage2_emfw}.

Both layers of the Stage-2 trigger have been installed, and all optical links have been validated.
Data transfers confirm correct mapping at each layer.
The Stage-2 trigger is currently running in parallel with Stage-1.
Collision data are being used to determine future calibrations and to study the performance of the algorithms.

\section{Conclusions}\label{sec:conclusions}

In Run 2, instantaneous luminosity and pile-up at the LHC will exceed design performance.  
The partial Stage-1 upgrade to the CMS L1 calorimeter trigger will maintain acceptance for proton and heavy ion collision events of interest during 2015 data-taking.
This is accomplished with event-by-event pile-up corrections in proton-proton collision running and a similar background subtraction in heavy ion collision running.
Starting in 2016, the Stage-2 trigger will go online.  Whole events are time multiplexed to data processors with full trigger tower granularity to achieve improved energy and position resolution.

\acknowledgments

We gratefully acknowledge the following sources of funding: CERN, CNRS/IN2P3/P2IOLabEx  (France), GSRT (Greece), STFC (United Kingdom), and DOE and NSF (U. S. A.).


\begin{thebibliography}{9}

\bibitem{lhc}
M. Lamont,
\emph{Status of the LHC},
\emph{J. Phys. Conf. Ser.} {\bf 455} (2013) 012001.

\bibitem{tdr}
CMS Collaboration,
\emph{CMS Technical Design Report for the Level-1 Trigger Upgrade},
\emph{CMS TDR} {\bf 012} (2013).

\bibitem{run1trig}
CMS Collaboration,
\emph{CMS TriDAS project: Technical Design Report; 1, the trigger systems},
CERN-LHCC-2000-038 (2000).

\bibitem{mp7}
K. Compton et al.,
\emph{The MP7 and CTP-6: multi-hundred Gbps processing boards for calorimeter trigger upgrades at CMS}
\jinst{7}{2012}{C12024}

\bibitem{tcc}
R. Alemany,
\emph{CMS ECAL Off-detector Electronics},
\emph{11th International Conference On Calorimetry In High Energy Physics},
29 Mar -- 2 Apr, 2004 Perugia, Italy, 181.

\bibitem{oSLB}
J. C. Da Silva, J. Varela, and P. Parracho,
\emph{The optical Synchronization and Link Board project, oSLB},
\jinst{8}{2013}{C02036}.

\bibitem{oRSC}
M. Baber et al.,
\emph{Development and testing of an upgrade to the CMS level-1 calorimeter trigger},
\jinst{9}{2014}{C01006}.

\bibitem{amc13}
E. Hazen et al.,
{The AMC13XG: a new generation clock/timing/DAQ module for CMS MicroTCA},
\jinst{8}{2013}{C12036}.

\bibitem{ipbus}
R.Frazier, G. Iles, D. Newbold, and A. Rose,
\emph{Software and firmware for controlling CMS trigger and readout hardware via gigabit Ethernet},
\emph{Physics Procedia} {\bf 37} (2013) 1892-1899.

\bibitem{hlt_tau}
CMS Collaboration,
\emph{CMS High Level Trigger}, 
CERN-LHCC-2007-021 (2007).

\bibitem{software}
A. Taurok et al.,
\emph{The central trigger control system of the CMS experiment at CERN},
\jinst{6}{2011}{P03004}

\bibitem{dqm}
L. Tuura, A. Meyer, I. Segoni, and G. Della Ricca,
\emph{CMS data quality monitoring: Systems and experiences},
\emph{J. Phys. Conf. Ser.} {\bf 219} (2010) 072020. 

\bibitem{ctp7}
A. Svetek et al., 
\emph{The Calorimeter Trigger Processor Card: The Next Generation of High Speed Algorithmic Data Processing at CMS}, 
these proceedings.

\bibitem{alex} 
A. Zabi et al., 
\emph{Triggering on electrons, jets and tau leptons with the CMS upgraded calorimeter trigger for the LHC RUN II}, these proceedings.


\end{thebibliography}
\end{document}